\documentclass[conference]{IEEEtran}
\ifCLASSINFOpdf
\else
\fi
\DeclareMathAlphabet{\mathpzc}{OT1}{pzc}{m}{it}  
\usepackage{mathrsfs}
\usepackage{amsthm}
\usepackage{makeidx}
\usepackage{examples}
\usepackage[pdftex]{graphicx}  
\usepackage{amssymb}
\usepackage{graphicx}
\usepackage{algorithm}
\usepackage{epstopdf}
\usepackage{color}
\usepackage{url}
\graphicspath{{../eps/}{../ps/}}
\usepackage{psfrag}    
\usepackage{algorithm}
\usepackage{algorithmic}
\usepackage{amsmath}
\usepackage{hyperref}
\font\msbm=msbm10 at 10pt
\newcommand{\ZZ}{\mbox{\msbm Z}}

\newcommand{\FF}{\mbox{\msbm F}}
\def \Z {{\ZZ}}

\def \F {{\FF}}

\newtheorem{theorem}{Theorem}

\newtheorem{remark}[theorem]{Remark}
\newtheorem{example}[theorem]{Example}
\newtheorem{definition}[theorem]{Definition}

\begin{document}
%
\title{On Heterogeneous Regenerating Codes and Capacity of Distributed Storage Systems}

\author{
\IEEEauthorblockN{Krishna Gopal Benerjee}
\IEEEauthorblockA{Laboratory of Natural Information Processing\\
Dhirubhai Ambani Institute of Information\\and Communication Technology\\
Gandhinagar, Gujarat, 382007 India\\
Email: krishna\_gopal@daiict.ac.in}
\and
\IEEEauthorblockN{Manish Kumar Gupta}
\IEEEauthorblockA{Laboratory of Natural Information Processing\\
Dhirubhai Ambani Institute of Information\\and Communication Technology\\
Gandhinagar, Gujarat, 382007 India\\
Email: mankg@computer.org}
}


%


\maketitle

\begin{abstract}
Heterogeneous Distributed Storage Systems (DSS)  are close to real world applications for data storage. Internet caching system and peer-to-peer storage clouds are the examples of such DSS. In this work, we calculate the capacity formula for such systems where each node store different number of packets and each having a different repair bandwidth (node can be repaired by contacting a specific set of nodes).  The tradeoff curve between storage and repair bandwidth is studied for such heterogeneous DSS. By analyzing the capacity formula new minimum bandwidth regenerating (MBR) and minimum storage regenerating (MBR) points are obtained on the curve. It is shown that in some cases  these are better than the homogeneous DSS.
\end{abstract}


%
\IEEEpeerreviewmaketitle

\section{Introduction}
Data storage has been a challenge for mankind since ancient times. Recently emerged Cloud computing provides an excellent way to store the data in a Distributed Storage Systems (DSSs).  Many such commercial 
systems are in use such as Hadoop based DSS of Facebook, Yahoo,  IBM, Amazon and Microsoft Windows Azure system \cite{XorbasVLDB,Huang:2012:ECW:2342821.2342823,skydrive,amazonec2}. In such a DSS, data is stored on $n$ nodes each of which may be unreliable hence data reliability is a major challenge for researchers. For example, in one month the maximum number of node failures is approximately 110 out of 3000 nodes in Facebook clusters \cite{XorbasVLDB}.  In the case of node failure, system has to repair the failed node by either generating functional equivalent of the data loss or by generating the exact data that was lost on that node. In order to provide reliability in these, systems either uses simple replication or they use  MDS (maximum distance separable) erasure codes. Simple replication uses more space (so it is bad for storage minimization) and erasuer MDS code approach is not efficient for bandwidth minimization in a node repair process. To optimize these conflicting parameters of data storage and bandwidth, recently in a seminal paper Dimakis et. al \cite{dgwr7} introduced regenerating codes and  later they were studied by many researchers ~\cite{XorbasVLDB,dgwr7, rr10,5961826,shah2012distributed, survey,dress11,DBLP:journals/corr/abs-1211-1932,2013arXiv1302.0744K,Gupta:arXiv1302.3681,2014arXiv1401.4734S,2014arXiv1401.4509A}.

Consider a DSS of total $n$ nodes. Regenerating codes are specified by the parameters  $\{[n, k, d], [\alpha,\beta, B] \}$, where B is the size of the file and $\alpha$ is the number of packets on each node.
In order to get a file user has to contact $k (k < n)$ nodes out of total $n$ nodes. In case of a node failure, data can be recovered by contacting $d$ nodes ($d$ is known as repair degree) and downloading $\beta$ packets from each node. Thus total bandwidth for a repair is $d\beta$. One has to optimize both $\alpha$ and $\beta$, hence we get two kind of regenerating codes viz. Minimum Storage Regenerating (MSR) codes useful for archival purpose and Minimum Bandwidth Regenerating (MBR) codes useful for Internet applications \cite{rr10,survey}. 
Most of the previous work in this area has been focused on homogeneous DSS and related regenerating codes. For example, using network flow analysis,  bounds on the capacity of DSS (the maximum amount of information delivered to any user contacting $k$ nodes out of $n$ nodes) are calculated for $(n,k,d)$ homogeneous DSS with symmetric repair  \cite{capacity}.  Using similar approach, in \cite{5513353}, Shah et al calculated cut-set lower bound on repair bandwidth for a special flexible setting  for homogeneous DSS.  Recently some work has been done for a more general setting where storage capacity of each node may vary. In particular,  in ~\cite{DBLP:journals/corr/abs-1211-0415},  Ernvall et al calculated the capacity bounds of a heterogeneous DSS having dynamic repair bandwidth and in \cite{6620424} a non-homogeneous two rack model of DSS has been considered.  Inspired by heterogeneous DSS new bounds and codes have been studied \cite{2014arXiv1402.2343G}. 

In the present work, we consider a heterogeneous DSS for which each storage node has different storage size and repair bandwidth. User can reconstruct the file by contacting any $k (< n$) nodes. In case of a node failure, data collector contacts specific set of helper nodes  and downloads fixed number of packets from each helper node. We calculated capacity formula for such a heterogeneous DSS.  Using this we obtain new MSR and MBR points on the tradeoff curve between storage and bandwidth.

\par \textit{Organization}: The paper is organized as follows. Section $2$ describes the model of our heterogeneous DSS and collects the necessary background. Main results of the paper are given in Section $3.$ Section $4$ gives a proof of Theorem $4$ and analysis of Theorem $6$ is given in Section $5$. Final section concludes the paper with general remarks.

 
\section{Model} 
In heterogeneous DSS a file is divided into encoded packets and they are distributed among $n$ nodes  $U_i \;(1 \leq i \leq n)$ each having storage capacity $\alpha_i \; (1 \leq i \leq n)$ and repair degree $d_i  \; (1 \leq i \leq n)$.
An user can reconstruct the file by downloading data from any $(k  < n)$ nodes.  If a node $U_i\;(1\leq i\leq n)$ fails then data collector will download $\beta$ packets from specific $d_i$ helping nodes. In such a case repair bandwidth for a node $U_i$ is $\gamma_i = d_i\beta$. We consider single node failure in our discussions.

\begin{definition}(Surviving Set): In a $(n,k)$ heterogeneous DSS, surviving set of a node $U_i\;(1\leq i\leq n)$ is a set of $d_i$ nodes which are used for repairing the node $U_i$. Note that there could be several surviving sets for a given node.  Indexing all the surviving sets by a positive integer $\ell,$ let us denote them by  $S_i^{(\ell)}$. 
\end{definition}

\begin{figure}
\centering
\includegraphics[scale=0.3]{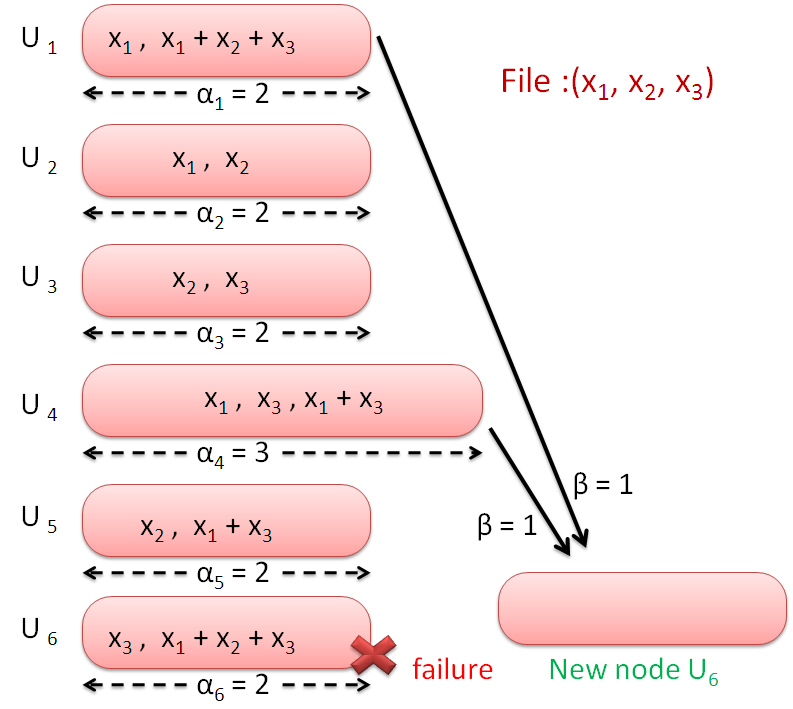}
\caption{A file is divided into $3$ distinct coded packets $x_1,x_2$ and $x_3$ from field $\F_q$. These three packets are encoded into five distinct packets and some copies of the five packets are distributed among $6$ nodes such a way that any data collector can download whole file by contacting any $2$ nodes. In this ($6,2$) heterogeneous DSS, repair degrees and the number of packets on each storage node $U_i\;\ (1\leq i\leq 6)$ are $2,2,2,3,2,2$ respectiveley . If a node $U_i$ fails then each helper node of any one surviving set $S^{(\ell)}_i$ of node $U_i$ will download $1$ packet to repair the  node $U_i\ i.e.\ \beta = 1$ unit.}
\label{example}
\end{figure}

\begin{example}
Consider a $(6, 2)$ heterogeneous DSS as shown in Figure (\ref{example}). The surviving sets of nodes are shown in table (\ref{example table}).
\begin{table}[ht]
\caption{Surviving sets of Heterogeneous DSS $(6, 2)$ shown in figure (\ref{example}).}
\centering 
\begin{tabular}{|c||c|}
\hline
Nodes&All possible surviving sets\\[0.5ex]
$U_i$& $S_i^{(\ell )}$                                                 \\
\hline\hline
$U_1$&$S_1^{(1)} = \{U_4, U_6\}, S_1^{(2)} = \{U_2, U_6\}.$            \\
\hline
$U_2$&$S_2^{(1)} = \{U_1, U_3\}, S_2^{(2)} =  \{U_1, U_5\},$            \\
     &$ S_2^{(3)} = \{U_4, U_3\},S_2^{(4)} =  \{U_4, U_5\}$.             \\
\hline 
$U_3$&$S_3^{(1)} = \{U_2, U_4\}, S_3^{(2)} =  \{U_2, U_6\},$            \\
     &$ S_3^{(3)} = \{U_5, U_4\},S_3^{(4)} =  \{U_5, U_6\}$.             \\
\hline 
$U_4$&$ S_4^{(1)} =  \{U_1, U_3, U_5\}, S_4^{(2)} =  \{U_2, U_3, U_5\},$\\
     &$ S_4^{(3)} = \{U_1, U_6, U_5\}, S_4^{(4)} = \{U_2, U_6, U_5\}.$ \\
\hline 
$U_5$&$ S_5^{(1)} =  \{U_2, U_4\}, S_5^{(2)} =  \{U_3, U_4\}.$\\
\hline
$U_6$&$ S_6^{(1)} =  \{U_3, U_1\}, S_6^{(2)} =  \{U_4, U_1\}.$          \\
\hline
\end{tabular}
\label{example table}
\end{table}
\end{example}

We are now in a position to describe the capacity of our heterogeneous DSS.  Using the information flow graph, the capacity of homogeneous DSS with symmetric repair was calculated in \cite{capacity} \begin{equation}\mathpzc{C}(\alpha, \gamma)=\sum_{i=1}^k\min\{\alpha,(d-i+1)\frac{\gamma}{d}\}.\end{equation} In our heterogeneous DSS, in case of a single node failure, it can be recovered by some specific set of the $d_i$ surviving  nodes. Thus a typical information flow graph representing a $(n, k)$ DSS is shown in Figure \ref{Use in lemma 1}.  A pair of graph node $In_i$ and $Out_i (1\leq i\leq n)$ in $G$, represents the storage node $U_i$. Here node $``s"$ is the source of whole file . If $\alpha _i$ is the storage capacity of node $U_i$ then the weight of directed edge $(In_i, Out_i)$ in flow graph is $\alpha _i$ because the node $U_i$ can flow $\alpha_i$ amount of information across the graph $G$.  If a node $U_i$ ($i.e.$ node pair $(In_i, Out_i)$) fails then all helping nodes of any one of the surviving set for storage node $U_i$ will download $\beta$ packets and generate a new storage node $U_i$ ($i.e.$ new node pair ($In_i', Out_i'$)). Now in order to calculate the maximum amount of packets that can be delivered to data collector (DC) by contacting any $k$ nodes (for $G$ any $k$ number of `out nodes' called `$Out_i$') one has to compute the min-cut of the information flow graph $G.$ Also to compute the cut in $G$ one requires a specific sequence of surviving sets picked up randomly one from each node. Formally we can define them as follows.
\begin{definition} (Surviving Sequence): For a $(n, k)$ heterogeneous DSS, surviving sequence $\left\langle\eta_j\right\rangle_{j=1}^n$ is a sequence of surviving sets picked up randomly one from each node.
\end{definition}
For example in $(6, 2)$ heterogeneous DSS as shown in Figure (\ref{example}) one of the possible surviving sequence $\left\langle \eta_j\right\rangle_{j=1}^{6}$ is $\left\langle S^{(1)}_5, S^{(1)}_3, S^{(2)}_4, S^{(2)}_6, S^{(1)}_1, S^{(2)}_2\right\rangle$. Another possible surviving sequence could be $\left\langle S^{(1)}_2, S^{(1)}_3, S^{(2)}_4, S^{(2)}_6, S^{(2)}_1, S^{(2)}_5\right\rangle$.
\section{Main Results}
In this section we describe our main results. The cut$(V_1,V_2)$ of a weighted graph $G(V,E)$ is the partition $V_1$ and $V_2$ ( where $V_1 \cap V_2 = \phi)$ of the vertex set $V$ such that at least one edge exist between $V_1$ and $V_2$. For source $s$ and sink $t$, the $s-t$ cut$(V_1,V_2)$ of a weighted graph is a cut such that  $s \in V_1$ and $t \in V_2$. For a graph $G(V,E)$, the ($s,t$)cut-set$(V_1,V_2)$ is the set of all edges between the partitions $V_1$ and $V_2$.
Hence for a weighted graph $G(V,E)$ min-cut($s,t$) is the sum of minimum possible weights of edges associated with ($s,t$)cut-set$(V_1,V_2)$ for arbitrary vertex set partitions $V_1$ and $V_2$.
Theorem (\ref{lemma 1}) gives the lower bound of min-cut between source node and data collector.  We also give new MSR and MBR points on storage and bandwidth tradeoff curve.
\begin{theorem}
For a $(n, k)$ heterogeneous DSS  the min-cut between source $s$ and data collector $t$ must satisfy the following inequality
\begin{equation}
\begin{split}
& \mbox{min-cut}(s,t)\geq 
\\ & \min_{\left\langle\eta_m\right\rangle_{m=1}^n\in\mathscr{T}}\left\{\sum\limits_{j=1}^{k}\min \left\{\alpha _j, \left|\eta_j\backslash \left(\bigcup_{\lambda=0}^{j-1}\{U_{\lambda}\}\right)\right|\beta \right\}\right\}, 
\end{split}
\label{lemma for mincut 1}
\end{equation}
where $\{U_0\}=\phi,\ 0\leq \lambda < j\leq k$, $\eta_j\in\left\langle\eta_m\right\rangle_{m=1}^n$ and $\mathscr{T}$ is the set of all possible surviving sequences $\left\langle  \eta_m \right\rangle_{m=1}^n$. 
\label{lemma 1}
\end{theorem}

It is easy to observe that for every $(n, k)$ heterogeneous DSS, there exist an information flow graph $G$ such that cut set achieves inequality (\ref{lemma for mincut 1}) with equality. But total information size delivered to data collector $t$ must be at-least equal to file size $B$ so the necessary condition for heterogeneous DSS is
\begin{equation}
 B\leq \min_{\left\langle\eta_m\right\rangle_{m=1}^n\in\mathscr{T}}\left\{\sum\limits_{j=1}^{k}\min \left\{\alpha _j, \left|\eta_j\backslash \left(\bigcup_{\lambda=0}^{j-1}\{U_{\lambda}\}\right)\right|\beta \right\}\right\},
\label{condition for B 1}
\end{equation}
By the definition of capacity $\mathpzc{C}$ of $(n, k)$ heterogeneous DSS one can define:
\begin{equation}
\mathpzc{C}\triangleq \min_{\left\langle\eta_m\right\rangle_{m=1}^n\in\mathscr{T}}\left\{\sum\limits_{j=1}^k\min \left\{\alpha _j, \left|\eta_j\backslash \left(\bigcup_{\lambda=0}^{j-1}\{U_{\lambda}\}\right)\right|\beta \right\}\right\}, 
\label{c 1}
\end{equation}

\begin{example}
For $(6, 2)$ DSS as shown in Figure (\ref{example}), the $capacity$ is $3$ units with $\beta=1$ unit.
\end{example}
\par 
One can determine the time complexity  for calculating the  capacity by equation (\ref{c 1}), with respect to the parameter of heterogeneous DSS. It is easy to observe that this time complexity depands on the number of possible surviving sequences. If there exist exactly one surviving set for every node then time complexity to calculate the capacity is $\Theta\left(\frac{n!}{(n-k)!}\right)$. In general, if there exist many surviving sets for a node then the time complexity for calculating  the capacity is $O\left(n!\prod_{i=1}^nd_i\right)$. This can be further improved to  $\Theta\left(k!\sum_{A\in\mathscr{A}}\prod_{i\in A}d_i\right)$, where $\mathscr{A}$ = $\left\{A:A\subseteq\{1,2,\ldots,n\},\left|A\right|=k\right\}$.
The new MSR and MBR points on the tradeoff curve of storage and bandwidth for our new heterogeneous DSS can be calculated with an approach similar to homogeneous DSS \cite{capacity}. The results are summarize in the following:
\begin{theorem}
For a surviving sequence $\left\langle  \eta_j^* \right\rangle_{j=1}^n\in \mathscr{T}$ that minimizes the RHS of inequality (\ref{lemma for mincut 1})  the new MSR and MBR points on the tradeoff curve between storage and bandwidth for heterogeneous DSS is given by the following equations: 
\begin{enumerate}
	\item For MSR point:
\begin{itemize}
	\item $\sum\limits_{j=1}^k\alpha_j=B$ (with $\alpha_1\leq\alpha_2\leq\ldots\leq\alpha_k$) and
	\item $\beta\leq\frac{B}{k}\left[\min \left\{\left|\eta_j^*\backslash \left(\bigcup_{l=0}^{j-1}\left\{U_l\right\}\right)\right|\right\}_{j=1}^k\right]^{-1}$.
\end{itemize}
	\item  For MBR point:
\begin{itemize}
	\item $\beta=B\left[\sum\limits_{j=1}^{k}\left|\eta_j^*\backslash \left(\bigcup_{l=0}^{j-1}\left\{U_l\right\}\right)\right|\right]^{-1}$
	\item $\alpha_i=Bd_i\left[\sum\limits_{j=1}^{k}\left|\eta_j^*\backslash \left(\bigcup_{l=0}^{j-1}\left\{U_l\right\}\right)\right|\right]^{-1}$.
\end{itemize}
\end{enumerate}
\label{MSRMBR}
\end{theorem}

\section{Proof  of Theorem (\ref{lemma 1})}
\begin{proof}
The proof is similar to the proof in homogeneous case \cite{capacity}.  Consider the information flow graph $G$ for a $(n, k)$ heterogeneous DSS as shown in Figure \ref{Use in lemma 1}. We focus on single node faliure. In order to calculate min-cut$(s,t)$ we  compute cut contribution by each node to min-cut$(s,t)$ (in case of single node failure) successively.  Suppose the cut of the information flow graph $G$ is $\left(V, V'\right)$. Observe that both vertex sets $V$ and $V'$ are non empty. Let $\mathcal{D}= min-cut (s,t)$ be the collection of cut edges $i.e.$  each edge in $\mathcal{D}$ joins a vertex from $V$ to a vertex from $V'$. For an arbitrary surviving sequence $\left\langle \eta_i\right\rangle_{i=1}^n$ = $\left\langle S^{(\ell_1)}_1, S^{(\ell_2)}_2, \ldots, S^{(\ell_n)}_n\right\rangle$ one can observe the following for each node failure:
\par WLOG,  we can assume if node $U_1$ fails then all helper nodes from $S^{(\ell_1)}_1=\eta_1$ will generate a new node. To calculate cut contribution for node $U_1$ one has to find the weight of $(in_1, Out_1)$ and sum of weight of all those edges which represent downloading $\beta$ amount of data to repair the node $U_1$ by $\eta_1$. Hence for $Out_1\in V'$ we can have the following cases:
\begin{enumerate}
	\item If $In_1 \in V$ then edge $(in_1, Out_1)\in \mathcal{D}$ or
	\item If $In_1 \in V'$ then all the $d_1$ number of edges associated with a specific $S^{(\ell_1)}_1$ are in $\mathcal{D}$ $i.e.$ for some $m\in\{2,3,..., n\}, (In_m, Out_m)\in \mathcal{D}$ where $U_m \in S^{(\ell_1)}_1=\eta_1$. 
\end{enumerate}
Hence the contribution for the cut$(s,t)$ of vertex $Out_1$ is $\min \left\{\alpha_1, \left| S^{(\ell_1)}_1\right|\beta\right\}$ = $\min \left\{\alpha_1, \left|\eta_1\backslash\left\{U_0\right\}\right|\beta\right\}$, where $\{U_0\}=\phi$. 
\par Similarly for the vertex $Out_2\in V'$ (associated with $\eta_2$),  the following two cases arises:
\begin{enumerate}
	\item If $In_2 \in V$ then edge $(in_2, Out_2)\in \mathcal{D}$ or
	\item If $In_2 \in V'$ then all the $\left| S^{(\ell_2)}_2\backslash \{U_1\}\right|$ number of edges associated with a spacific $S^{(\ell_2)}_2$ are in $\mathcal{D}$ $i.e.$ for some $m\in\{3,4,..., n\}, (In_m, Out_m)\in \mathcal{D}$, where $U_m \in S^{(\ell_2)}_2=\eta_2$ and $\left| S^{(\ell_2)}_2\backslash \{U_1\}\right|$ distinct values are possible for $m$.
\end{enumerate}
So the contribution for the cut$(s,t)$ of vertex $Out_2$ is $\min \left\{\alpha_2, \left|S^{(\ell_2)}_2\backslash \left\{U_1\right\}\right|\beta\right\}$ $i.e.$ $\min \left\{\alpha_2, \left|\eta_2\backslash \left\{U_1\right\}\right|\beta\right\}$.
\par Similarly for $Out_3$ (associated with $\eta_3$)  in $V'$,  the two cases are:
\begin{enumerate}
	\item If $In_3 \in V$ then edge $(in_3, Out_3)\in \mathcal{D}$ or
	\item If $In_3 \in V'$ then all the $\left|S^{(\ell_3)}_3\backslash\left(\left\{U_1\right\}\cup \left\{U_2\right\}\right)\right|$ number of edges associated with a spacific $S^{(\ell_2)}_2$ are in $\mathcal{D}$ $i.e.$ for some $m\in\{4,5,..., n\}, (In_m, \\ Out_m)\in \mathcal{D}$, where $U_m \in S^{(\ell_3)}_3\backslash\left(\left\{U_1\right\}\cup \left\{U_2\right\}\right)\subset S^{(\ell_3)}_3$, $S^{(\ell_3)}_3=\eta_3$ and $\left|S^{(\ell_3)}_3\backslash\left(\left\{U_1\right\}\cup \left\{U_2\right\}\right)\right|$ distinct values are possible for $m$.
\end{enumerate}
So the contribution for the cut$(s,t)$ of vertex $Out_3$ is \newline $\min\{\alpha_2, \left|S^{(\ell_3)}_3\backslash\left\{U_1, U_2\right\}\right|\beta\} = \min \left\{\alpha_2, \left|\eta_3\backslash \left\{U_1, U_2\right\}\right|\beta\right\}$.
\par Continuing in the same way  for a vertex $Out_j \in V'$ (associated with $\eta_j$) the following two cases are possible:
\begin{enumerate}
	\item If $In_j \in V$ then edge $(in_j, Out_j)\in \mathcal{D}$ or
	\item If $In_j \in V'$ then all the $\left|S^{(\ell_j)}_j\backslash \left(\bigcup_{l=0}^{j-1}\left\{U_l\right\}\right)\right|$ number of edges associated with a spacific $S^{(\ell_j)}_j$ are in $\mathcal{D}$ $i.e.$ for some $m\in\{j+1,j+2,..., n\}, (In_m, \\ Out_m)\in \mathcal{D}$ where $U_m \in S^{(\ell_j)}_j\backslash\left(\bigcup_{l=0}^{j-1}\left\{U_l\right\}\right)\subset S^{(\ell_j)}_j$, $S^{(\ell_j)}_j=\eta_j$ and $\left|S^{(\ell_j)}_j\backslash\left(\bigcup_{l=0}^{j-1}\left\{U_l\right\}\right)\right|$ distinct values are possible for $m$.
\end{enumerate}
So the contribution for the cut$(s,t)$ of the vertex $ Out_j$ is $\min \left\{\alpha _j, \left|S^{(\ell_j)}_j\backslash \left(\bigcup_{l=0}^{j-1}\left\{U_l\right\}\right)\right|\beta \right\}$\newline = $\min \left\{\alpha _j, \left|\eta_j\backslash \left(\bigcup_{l=0}^{j-1}\left\{U_l\right\}\right)\right|\beta \right\}$, where $\eta_j=S^{(\ell)}_j$ and $\eta_j\in\left\langle \eta_j\right\rangle_{j=1}^n$. Thus min-cut$(s, t)$ for graph $G$ will satisfy

\begin{equation*}
\begin{split}
& \mbox{min-cut}(s,t)\geq 
\\ & \min_{\left\langle\eta_m\right\rangle_{m=1}^n\in\mathscr{T}}\left\{\sum\limits_{j=1}^{k}\min \left\{\alpha _j, \left|\eta_j\backslash \left(\bigcup_{\lambda=0}^{j-1}\{U_{\lambda}\}\right)\right|\beta \right\}\right\}, 
\end{split}
\end{equation*}
where $\{U_0\}=\phi,\ 0\leq \lambda < j\leq k$, $\eta_j\in\left\langle\eta_m\right\rangle_{m=1}^n$ and $\mathscr{T}$ is the set of all possible surviving sequences $\left\langle  \eta_m \right\rangle_{m=1}^n$. 

\begin{figure}
\centering
\includegraphics[scale=0.26]{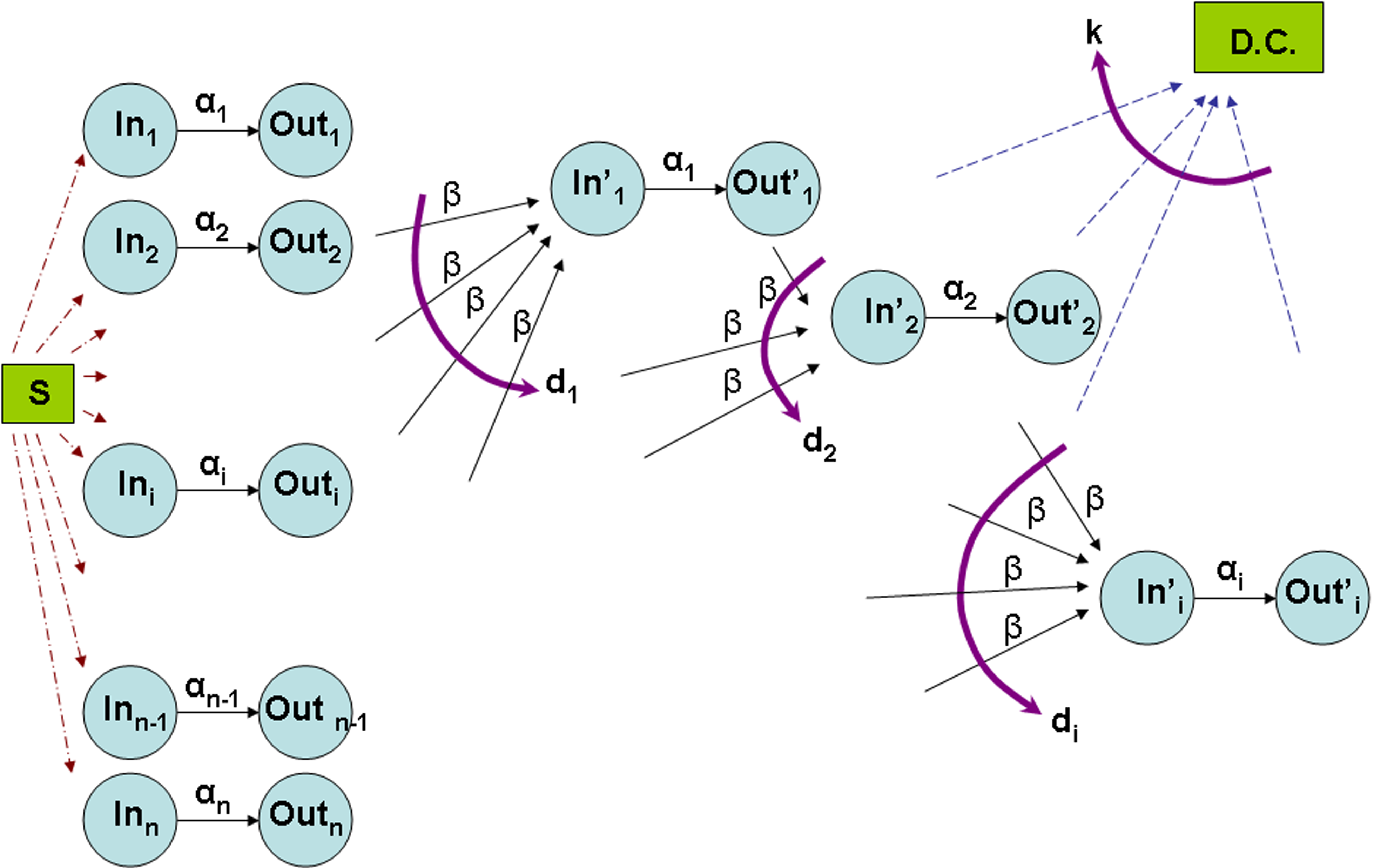}
\caption{As per assumptions, ($n,k$) heterogeneous DSS model can tolerate only one node failure at a time. WLOG one can assume node $U_1$ fails and then repairs itself using  helping nodes of surviving set $S^{(\ell_1)}_1$ after that node $U_2$ fails and repairs itself using surviving set $S^{(\ell_2)}_2$ and so on. Suppose after some time node $U_i$ fails and then repairs by contacting surviving set $S^{(\ell_i)}_i$. Then $\min\{\alpha_1,\left| S^{(\ell_1)}_1\right|\}$, $\min\{\alpha_2,\left| S^{(\ell_2)}_2\backslash \{U_1\}\right|\}$, \ldots , $\min\{\alpha_i,\left| S^{(\ell_i)}_i\backslash \{U_1,U_2, \ldots, U_{i-1}\}\right|\}$ number of times of downloading of $\beta$ amount of information is required to make DSS stable.
}
\label{Use in lemma 1}
\end{figure}
\end{proof}

\section{Analysis of Theorem \ref{MSRMBR}}
In heterogeneous DSS the RHS of min-cut inequality (\ref{lemma for mincut 1}) is maximum possible amount of data that can be delivered to any data collector by contacting any $k$ nodes. By the relation (\ref{condition for B 1}) one can draw optimal tradeoff for heterogeneous DSS.

\begin{figure}
\centering
\includegraphics[scale=0.35]{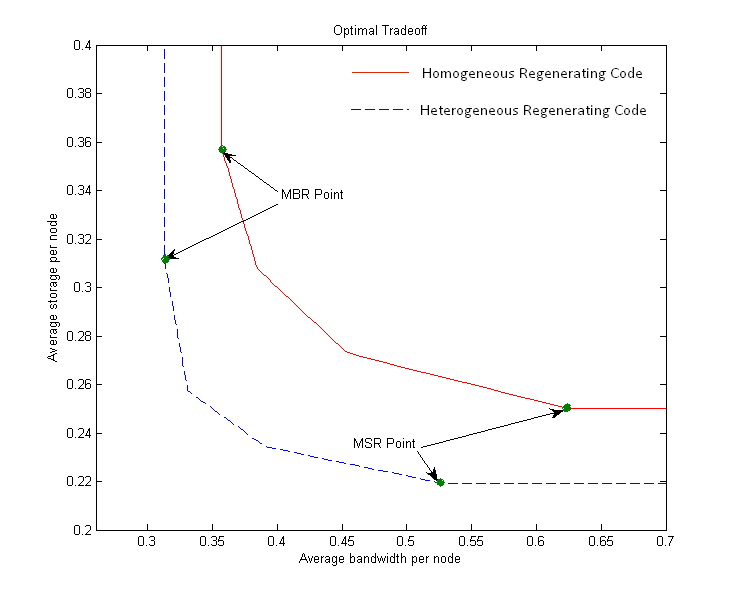}
\caption{Comparative analysis of optimal tradeoff between ($7,4$) heterogeneous DSS and ($7,4,5$) homogeneous DSS. For heterogeneous DSS file size B = $1$ unit, reconstruction degree $k$ = 4 nodes and total numbers of nodes $n$ = 7. The tradeoff is drown for the heterogeneous DSS having repair degree $d_i$ and storage capacity $\alpha_i$ are proportional to $4:4:5:5:5:6:6$ for $i\in\{1, 2, 3, 4, 5, 6, 7\}$.} 
\label{tradeoff}
\end{figure}

By the inequality (\ref{condition for B 1}) one can conclude that 
\begin{equation}
    \exists i (1\leq i\leq n)\;\mbox{such that}\; \min \left\{\left|\eta_j^*\backslash \left(\bigcup_{l=0}^{j-1}\left\{U_l\right\}\right)\right|\beta\right\}_{j=1}^k\leq\alpha_i
\label{inequality 1}
\end{equation}
For if $\min \left\{\left|\eta_j^*\backslash \left(\bigcup_{l=0}^{j-1}\left\{U_l\right\}\right)\right|\beta\right\}_{j=1}^k>\alpha_i$ where $ (1\leq i \leq n)$ then one can reduce $\beta$ without violating the inequality (\ref{condition for B 1}).
On the other hand to repair a failed node $U_i$ one has to download at least $\alpha_i$ packets that are stored in node $U_i$ so 
\begin{equation}
\alpha_i \leq d_i\beta.
\label{bandwidth 1}
\end{equation}

\par Hence by (\ref{inequality 1}) and (\ref{bandwidth 1}),
\[
    \min \left\{\left|\eta_j^*\backslash \left(\bigcup_{l=0}^{j-1}\left\{U_l\right\}\right)\right|\beta\right\}_{j=1}^k\leq\alpha_i \leq d_i\beta.
\]
Thus for reconstructing the file by contacting any $k$ nodes we have:
\begin{equation}
k\min_{1\leq j\leq k} \left\{\left|\eta_j^*\backslash \left(\bigcup_{l=0}^{j-1}\left\{U_l\right\}\right)\right|\beta\right\}\leq\sum\limits_{m\in A}\alpha_m \leq \beta\sum\limits_{m\in A}d_m,
\label{alpha}
\end{equation}
where an arbitrary $A\in\mathscr{A}= \{A:A\subseteq\{1,2,...,n\}, \left|A\right| =k\}$.
\par For calculating MSR point, first minimize storage and then bandwidth to get:    
\begin{itemize}
	\item $\sum\limits_{j=1}^k\alpha_j=B$ (with $\alpha_1\leq\alpha_2\leq\ldots\leq\alpha_k$) and
	\item $\beta\leq\frac{B}{k}\left[\min \left\{\left|\eta_j^*\backslash \left(\bigcup_{l=0}^{j-1}\left\{U_l\right\}\right)\right|\right\}_{j=1}^k\right]^{-1}$.
\end{itemize}
\par Similarly for calculating MBR point first minimize bandwidth and then storage. Hence inequality (\ref{condition for B 1}) reduces into
\[
\alpha _i\geq \max\left\{\left|\eta_j^*\backslash \left(\bigcup_{l=0}^{j-1}\left\{U_l\right\}\right)\right|\beta\right\}_{j=1}^k\ \ \forall i
\]
Hence for MBR point
\begin{itemize}
	\item $\beta=B\left[\sum\limits_{j=1}^{k}\left|\eta_j^*\backslash \left(\bigcup_{l=0}^{j-1}\left\{U_l\right\}\right)\right|\right]^{-1}$
	\item $\alpha_i=Bd_i\left[\sum\limits_{j=1}^{k}\left|\eta_j^*\backslash \left(\bigcup_{l=0}^{j-1}\left\{U_l\right\}\right)\right|\right]^{-1}$
\end{itemize}
\begin{remark}
Note that the MSR and MBR points of \cite{5550492} for homogeneous DSS  are special case of new points. In particular, 
for an arbitrary $\ell$, if 
\begin{enumerate}
	\item $\alpha_i=\alpha$, $ \left|S^{(\ell)}_i\right|=d_i= d,\ \forall i\in\{1,2,...,n\}\ $ and
	\item a failed node $U_i$ can be repair by any $d$ nodes in system
\end{enumerate}
then the parameter $\alpha$ and $\beta$ of
\begin{enumerate}
	\item MSR point would be
	\[
	\alpha\ =\ \frac{B}{k},\ \beta\ =\ \frac{B}{k(d-k+1)}
	\]
	\item MBR point would be
	\[
	\alpha\ =\ \frac{2Bd}{k(2d-k+1)},\ \beta\ =\ \frac{2B}{k(2d-k+1)}.
	\]
\end{enumerate}
\end{remark}
\begin{example} For $(6, 2)$ heterogeneous DSS as shown in Figure (\ref{example}) with file size $B=3$, one can calculate $\alpha_i$ and $\beta$ for MBR and MSR points. For MSR point $\alpha_j=3/2$, $\alpha_4=1$ and $\beta=3/2$, where $j\in\{1,2,3,5,6\}$. Similarly for MBR point $\beta=1$ and $\alpha_j=2$ and  $\alpha_4=3$, where $j\in\{1,2,3,5,6\}$.
\end{example}
\section{Conclusion}
Motivated by  real world applications which are based on heterogeneous DSS with dynamic repair bandwidth and constant repair degree, we calculated capacity of heterogeneous DSS with dynamic repair bandwidth and dynamic repair degree (but with constant $\beta$). By analyzing the capacity formula new MSR and MBR points are obtained. We also show that in some cases the new MSR and MBR points are better than homogeneous DSS. It would be an interesting future task to construct efficient codes that meet these points. Also, in future, the capacity analysis could be generalized for heterogeneous DSS with dynamic repair degree and dynamic downloading factor.  
\bibliographystyle{IEEEtran}
\bibliography{cloud}
\end{document}